\documentclass[a4paper,11pt]{article}
\usepackage{pos}
\usepackage{comment}
\definecolor{ao(english)}{rgb}{0.0, 0.5, 0.0}
\renewcommand{\logo}{\relax}

\title{Pushing the Heavy Quark Expansion for $b\to cl\bar{\nu}$ to Higher Order in $1/m_b$}
\author*[a]{Ilija S.~Milutin}
\author[a]{Thomas Mannel}
\author[b,c]{K.~Keri Vos}

\affiliation[a]{Theoretische Physik 1, Center for Particle Physics Siegen \\
   Universit\"at Siegen,  D-57068 Siegen, Germany}

\affiliation[b]{Gravitational Waves and Fundamental Physics (GWFP), \\
Maastricht University, Duboisdomein 30, NL-6229 GT Maastricht, the Netherlands}

\affiliation[c]{Nikhef, Science Park 105, NL-1098 XG Amsterdam, the Netherlands}

\emailAdd{ilija.milutin@uni-siegen.de}
\emailAdd{Mannel@physik.uni-Siegen.de}
\emailAdd{k.vos@maastrichtuniversity.nl}

\abstract{The Heavy Quark Expansion (HQE) is the major tool to perform calculations for inclusive semileptonic $B\to X_cl\bar{\nu}$ and, consequently, for precision determinations of the CKM matrix element $V_{cb}$. To further improve precision, we pushed the expansion to $1/m_b^5$ to include even higher order terms in the HQE. Notably, at $1/m_b^5$, ``intrinsic charm'' (IC) contributions proportional to $1/(m_b^3m_c^2)$ occur, which are numerically expected to be sizeable. We will therefore firstly discuss the determination of a reduced set of HQE parameters at $1/m_b^5$, employing Reparametrisation Invariance (RPI) to this end. Consequently, we will show how the IC and ``genuine" $1/m_b^5$ contribute to the different kinematical moments of $B\to X_cl\bar{\nu}$, focusing in this work on the lepton energy $E_l$ moments. Using the ``lowest-lying state saturation ansatz'' (LLSA), we estimate the size of these contributions and observe a partial cancellation between the IC and ``genuine'' $1/m_b^5$ contributions, leading to an overall small contribution.

}

\FullConference{42nd International Conference on High Energy Physics (ICHEP2024)\\
18-24 July 2024\\
Prague, Czech Republic\\}

\renewcommand{\hookAfterAbstract}{%
\par\bigskip
}
\begin{document}
\maketitle

\section{Introduction}
\vspace*{-4mm}
\noindent 
The Heavy Quark Expansion (HQE) is the major tool to describe inclusive semileptonic $B\to X_c l\bar{\nu}$ decays, which in turn can be used to for a precision determination of the CKM element $|V_{cb}|$. This has resulted in determinations of $|V_{cb}|$ with a relative precision of about $1-2\%$ \cite{Bordone:2021oof,Bernlochner:2022ucr,Finauri:2023kte}. Increasing the precision of these determinations of $|V_{cb}|$ from inclusive decays, where one sums over final states in the decay, are relevant for resolving the puzzling tension between inclusive and exclusive (where one final state is chosen) $|V_{cb}|$ determinations (see e.g.~\cite{Gambino:2020jvv}).

The HQE is an expansion in $\Lambda_{\rm{QCD}}/m_Q$, where $m_Q$ is the mass of the heavy quark and $\Lambda_{\rm{QCD}}$ corresponds to the scale induced the running of the QCD coupling constant. The HQE for inclusive $b\to cl\bar{\nu}$ is set up as an Operator Product Expansion (OPE), employing full QCD heavy quark states. It manifests as a sum of HQE parameters, which are forward matrix elements of local operators of dimension $n$ containing the non-perturbative physics of order $\Lambda_{\rm{QCD}}^{n-3}$, which are in turn multiplied by coefficients which can be determined from perturbation theory as an expansion in $\alpha_s$ (as a function of the mass ratio $\rho\equiv m_c^2/m_b^2$). Therefore, the HQE is a double expansion in $\Lambda_{\rm{QCD}}/m_b$ and $\alpha_s$. At both order $\Lambda_{\rm{QCD}}^2/m_b^2$ and $\Lambda_{\rm{QCD}}^3/m_b^3$, two HQE parameters enter: $\mu_\pi^2,\, \mu_G^2$ and $\rho_D^3,\, \rho_{LS}^3$ respectively. However, at $\Lambda_{\rm{QCD}}^4/m_b^4$ the number of HQE parameters starts to proliferate, with 9 parameters at $\Lambda_{\rm{QCD}}^4/m_b^4$ and 18 at $\Lambda_{\rm{QCD}}^5/m_b^5$ \cite{Mannel:2010wj}.

Notably, the dilepton invariant mass $q^2$ moments recently measured by Belle \cite{Belle:2021idw} and Belle II \cite{Belle-II:2022evt}, are Reparamerization Invariant (RPI), which depend only on a reduced set of HQE parameters. This reduction of HQE parameters has allowed the extraction of $1/m_b^4$ HQE parameters and $|V_{cb}^{\rm{incl}}|$ at $\mathcal{O}(1/m_b^4)$ from data \cite{Bernlochner:2022ucr}. At $1/m_b^5$, inverse powers of the mass ratio $\rho$ appear for the first time, resulting in terms like $1/m_b^5\times 1/\rho=1/(m_b^3m_c^2)$, known as Intrinsic Charm (IC) \cite{Bigi:2009ym, Breidenbach:2008ua, Fael:2019umf}. Numerically, we have $m_c^2\approx m_b\Lambda_{\rm{QCD}}$, suggesting to count $\Lambda_{\rm{QCD}}^5/(m_b^3m_c^2)\sim \Lambda_{\rm{QCD}}^4/m_b^4$. Consequently, this suggests that a full analysis of $1/m_b^4$ needs to include these IC terms, based on power-counting arguments. Following \cite{Mannel:2023yqf}, we will discuss these power-enhanced IC terms, along with the other ``genuine'' $1/m_b^5$ contributions in the different kinematic moments of $B\to X_c l\bar{\nu}$.

\section{Setting up the Heavy Quark Expansion}
\vspace*{-4mm}
\noindent
We will consider the semileptonic inclusive decay $B(p_B)\to X_c(p_X)l(p_l)\bar{\nu}(p_\nu)$, where $q\equiv p_l+p_\nu$. The HQE is set up by employing the optical theorem to the correlation function of two weak $b\to c$ currents
\begin{align}
    R_{\mu\nu}(q)=\langle B(v)|\mathcal{R}_{\mu\nu}|B(v)\rangle=\int\text{d}^4x\, e^{iq\cdot x}\langle B(v)|T[\bar{b}(x)\Gamma_\mu c(x)\bar{c}(0)\bar{\Gamma}_\nu b(0)]|B(v)\rangle\ ,
\end{align}
where $\Gamma_\mu=\gamma_\mu(1-\gamma_5)$ and $|B(v)\rangle$ is the full QCD $B$ meson state moving at velocity $v=p_B/m_B$. Consequently, the heavy quark momentum is split according to $p_b=m_bv+k$, where $k$ is a small residual momentum of order $\Lambda_{\rm{QCD}}$. This corresponds to a field redefinition of the heavy quark according to $b(x)=e^{-im_b(v\cdot x)}b_v(x)$. Expanding in $k/m_b\sim iD/m_b$ will result in the OPE for $R_{\mu\nu}$, which can be related to the differential decay width $\text{d}\Gamma$. Schematically, this will result in
\begin{align}
    \text{d}\Gamma=\text{d}\Gamma^{(3)}+\frac{1}{m_b^2}\text{d}\Gamma^{(5)}+\frac{1}{m_b^3}\text{d}\Gamma^{(6)}+\frac{1}{m_b^4}\text{d}\Gamma^{(7)}+...\,,\quad\quad \text{d}\Gamma^{(n)}=\sum\limits_i \mathcal{C}_i^{(n)}\langle B|\mathcal{O}^{(n)}|B\rangle\ ,
\end{align}
where the $1/m_b$ term vanishes due to Heavy Quark Symmetries. The perturbative Wilson coefficients $\mathcal{C}_i^{(n)}$ are series in $\alpha_s$, which are multiplied by non-perturbative forward matrix elements $\langle B|\mathcal{O}^{(n)}|B\rangle$ of local operators $\mathcal{O}^{(n)}$ of dimension $n$, consisting of chains of covariant derivatives $iD$. For example, $\text{d}\Gamma^{(5)}$ contains $\mu_\pi^2$ and $\mu_G^2$, and $\text{d}\Gamma^{(6)}$ contains $\rho_D^3$ and $\rho_{LS}^3$, which are defined as\footnote{For the definitions of the other HQE parameters, we refer to \cite{Mannel:2023yqf}.} 
\begin{align}
    2m_B\mu_G^2=\langle B|\bar{b}_v(-\sigma^{\mu\nu})(iD_\mu)(iD_\nu)b_v|B\rangle\ ,\quad 2m_B\rho_D^3=\frac{1}{2}\langle B|\bar{b}_v[iD_\mu,[ivD,iD^\mu]]b_v|B\rangle\ .
\end{align}
Starting at $1/m_b^4$, the number of HQE parameters starts to proliferate with 9 new parameters in $\text{d}\Gamma^{(7)}$ and 18 in $\text{d}\Gamma^{(8)}$ (at tree level) \cite{Mannel:2010wj, Kobach:2017xkw}.
Since the HQE cannot predict the spectrum point-by-point, we study integrated quantities such as the kinematical moments, which are defined as
\begin{align}
    \langle (O)^n\rangle_{\rm{cut}}\rangle=\int_{\rm{cut}}(O)^n\frac{\text{d}\Gamma}{\text{d}O}\text{d}O\Bigg/\int_{\rm{cut}}\frac{\text{d}\Gamma}{\text{d}O}\text{d}O\ ,\label{eq:moment}
\end{align}
where $O$ are kinematical quantities $O\in\{q^2=(p_l+p_\nu)^2,\, E_l=v\cdot p_l,\, M_X^2=(m_Bv-q)^2\}$. The subscript "cut" indicates a restriction in the lower bound of the integration, required due to the experimental setup. We will focus on the centralized moments for $q^2$ and $E_l$ defined as
\begin{align}
    q_1(q^2_{\rm{cut}})&=\langle q^2\rangle_{q^2\geq q^2_{\rm{cut}}},\quad \quad q_n(q^2_{\rm{cut}})=\langle (q^2-\langle q^2\rangle)^n\rangle_{q^2\geq q^2_{\rm{cut}}}\ \ \text{for}\ n\geq2\ ,\\
    \ell_1(E_l^{\rm{cut}})&=\langle E_l\rangle_{E_l\geq E_l^{\rm{cut}}},\quad \quad \ell_n(E_l^{\rm{cut}})=\langle (E_l-\langle E_l\rangle)^n\rangle_{E_l\geq E_l^{\rm{cut}}}\ \ \text{for}\ n\geq2\ .
\end{align}
\section{Reparametrisation Invariance}
\vspace*{-4mm}
\noindent
As mentioned before, the number of HQE parameters starts to proliferate at $1/m_b^4$. However, the choice of $v_\mu$ in the splitting of the heavy quark momentum is not unique. This leads to the Reparametrisation Invariance (RPI) of the OPE and HQE \cite{Mannel:2018mqv, Luke:1992cs, Manohar:2010sf, Dugan:1991ak, Chen:1993np}. This can be imposed under the infinitesimal change $\delta_{\rm{RP}}: v_\mu\to v_\mu+\delta v_\mu$ and simultaneously $\delta_{\rm{RP}}iD_\mu=-m_b\delta v_\mu$ with $v\cdot \delta v=0$, which should leave $\langle B(v)|\mathcal{R}(S)|B(v)\rangle$ invariant (where $S=v-q/m_b$). As a result, RPI will link different orders in $1/m_b$ through \cite{Mannel:2018mqv}
\begin{align}
\delta_{\rm{RP}}C_{\mu_1...\mu_n}^{(n)}(S)=m_b\,\delta v^\alpha\left(C_{\alpha\mu_1...\mu_n}^{(n+1)}(S)+C_{\mu_1\alpha...\mu_n}^{(n+1)}(S)+...+C_{\mu_1...\mu_n\alpha}^{(n+1)}(S)\right)\ ,
\end{align}
which allows one to find combinations of HQE operators which are RPI, leading to a reduction of independent parameters for RPI quantities. In \cite{Mannel:2018mqv}, it has been worked out that up to $1/m_b^4$ one finds 8 independent RPI parameters, contrasting to the 13 HQE parameters in the full (non-RPI) basis. In \cite{Mannel:2023yqf}, this has been extended to $1/m_b^5$, finding 10 RPI parameters at this order, instead of 18 in the full non-RPI basis.

\subsection{\boldmath $|V_{cb}|$ from $q^2$ moments\unboldmath}
\vspace*{-2mm}
\noindent 
This reduction of HQE parameters due to RPI has been employed to extract inclusive $|V_{cb}|$ up to $1/m_b^4$ in \cite{Bernlochner:2022ucr}. The HQE parameters up to $1/m_b^4$ were extracted from the $q^2$ moments, since they are RPI, because they are independent of $v_\mu$ (as can be seen under \eqref{eq:moment}). These HQE parameters can then be combined with the measurements of $\mathcal{B}(B\to X_c l\bar{\nu})$ and the theoretical expression for total rate, which depends on the RPI HQE parameters. Consequently, $|V_{cb}^{\rm{incl}}|=(41.69\pm0.63)\times 10^{-3}$ has been extracted, resulting in the first extraction at $\mathcal{O}(1/m_b^4)$ and agreement at $1-2\sigma$ level with previous $\mathcal{O}(1/m_b^3)$ determinations \cite{Finauri:2023kte, Bordone:2021oof, Alberti:2014yda, Gambino:2013rza}. For a recent overview of the references of all known $1/m_b$ and $\alpha_s$ contributions of the total rate and kinematical moments, we refer to \cite{Fael:2024fkt,Fael:2024rys}.

\section{Intrinsic Charm}
\begin{figure}[t!]
    \centering
    \includegraphics[width=0.5\linewidth]{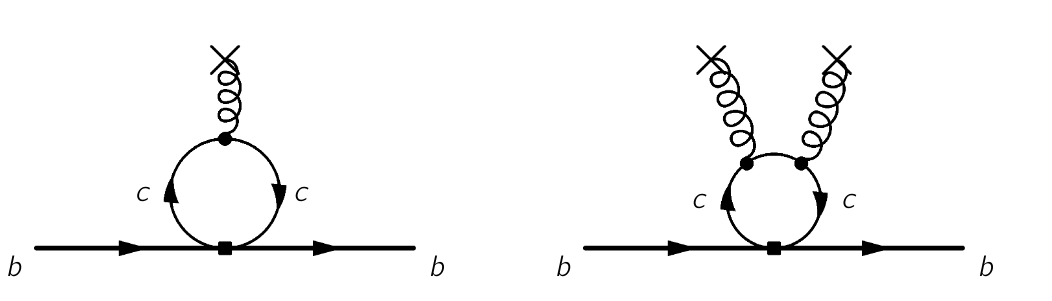}
    \caption{The Feynman diagrams of the intrinsic charm (IC) contributions.}
    \label{fig:FeynmanIC}
\end{figure}

\vspace*{-5.25mm}
\noindent
As already mentioned, at $1/m_b^5$, we find power-enhanced contributions of the form $1/m_b^5\times 1/\rho=1/(m_b^3m_b^2)$, known as Intrinsic Charm (IC) contributions, arising from diagrams like in figure \ref{fig:FeynmanIC}. Actually, since the charm and bottom quark are integrated out simultaneously, the HQE exhibits an infrared sensitivity to the charm mass starting at $1/m_b^3$ in the form of terms proportional to $\log{m_c^2/m_b^2}$. At higher dimensions $n$, power-like singularities for $m_c\to0$ occur proportional to $1/(m_b^3m_c^{n-6})$ where $n=8,10,12,...$. Including $\alpha_s$ corrections will introduce odd powers of $1/m_c$, however, we will focus on the tree-level contributions in this work.

Numerically, we may argue that $\Lambda_{\rm{QCD}}^5/(m_b^3 m_c^2)$ counts like $\Lambda_{\rm{QCD}}^4/m_b^4$, suggesting we require the IC contributions at dimension 8 to complete the analysis at $1/m_b^4$.

Therefore, we determine the full dimension-8 contribution and split of the IC contributions proportional to $1/\rho$ from the other ``genuine'' $1/m_b^5$ contributions at dimension 8. Interestingly, we find that we only need one combination of dimension-8 RPI operators, which we will denote by $X_{\rm{IC}}^5$, to describe the IC contributions in the total rate $\Gamma$ and $q^2$ moments. Furthermore, we find that the same combination $X_{\rm{IC}}^5$ also describes the IC contributions in the $E_l$ and $M_X^2$ moments.

Before we discuss the phenomenological implications, we require estimates for the size of the dimension-8 HQE parameters. For this, we employ the ``lowest-lying state saturation ansatz" (LLSA) \cite{Heinonen:2014dxa}, which provides estimates for higher dimension HQE parameters, by linking them to lower dimensional ones like $\mu_\pi^2,\ \mu_G^2,\ \rho_D^3,\ \rho_{LS}^3$ which have been determined from data. However, the LLSA unfortunately does not provide a way to estimate the uncertainty associated to the approximation, which is why we do not include uncertainties in the following. The input values used in the following can be found in \cite{Mannel:2023yqf}.

\begin{figure}[t!]
    \centering
    \includegraphics[width=0.46\linewidth]{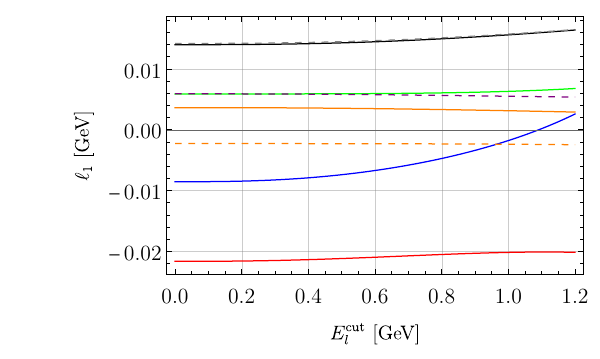}
    \includegraphics[width=0.46\linewidth]{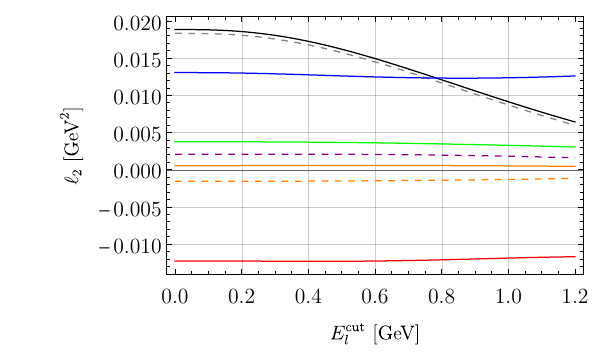}
    \includegraphics[width=0.46\linewidth]{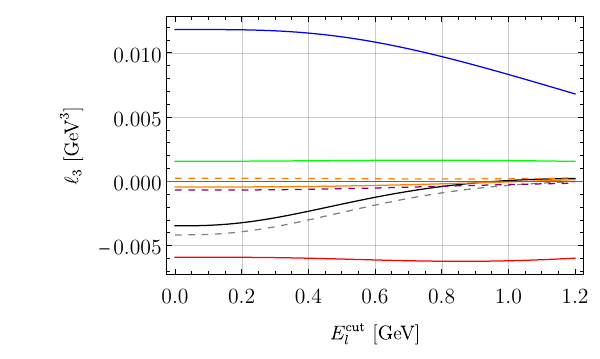}
    \includegraphics[width=0.46\linewidth]{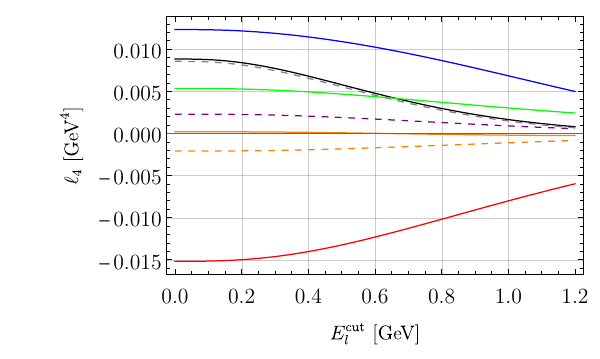}
    \includegraphics[width=0.95\linewidth]{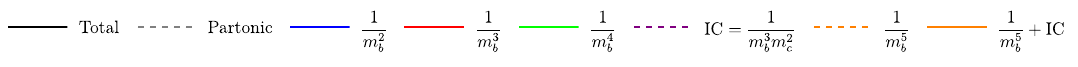}
    \caption{The first four lepton energy moments $\ell_n$ as a function of the lepton energy cut $E_l^{\rm{cut}}$. The total (solid black) and partonic (dashed gray) contribution of $\ell_1$ has been multiplied by 1/100 and of $\ell_{2,3,4}$ by 1/10, to show the contributions of the other orders more clearly. The solid blue, red, and green lines represent the $1/m_b^{2,3,4}$ contributions respectively. The dashed purple and orange lines represent the intrinsic charm and `genuine' $1/m_b^5$ contributions respectively, while the solid orange lines represent the sum of the latter two.}
    \label{fig:El_IC}
\end{figure}

In figure \ref{fig:El_IC}, we show the different $1/m_b$ contributions in the first four centralized $E_l$ moments, as a function of the lepton energy cut\footnote{A similar figure for the $q^2$ moments can be found in \cite{Mannel:2023yqf}.}. Within the LLSA estimates, we find that indeed the IC and ``genuine'' $1/m_b^5$ terms are roughly equal in size, but contribute with different signs in the $E_l$ (and $q^2$) moments. For the $q^2$ moments, these IC contributions can even be larger than the $1/m_b^4$ contributions, but are partially cancelled (just as in the $E_l$ moments) leading to an overall small contribution of the dimension-8 operators. Therefore, we recommend to only consider the IC terms in a combined determination up to $1/m_b^5$. For the $M_X^2$ moments, we find that the IC and ``genuine'' $1/m_b^5$ do contribute with the same sign, but that the IC contributions are very small and the total dimension-8 contribution is still smaller than the $1/m_b^4$ contributions (within the LLSA). Finally, by employing the LLSA estimates we can write the total rate for $B\to X_c l\bar{\nu}$ as
\begin{align}
    \Gamma=&\ \left(\frac{G_F^2m_b^5|V_{cb}|^2}{192\pi^3}\right)\Big(0.65\big|_{\mu_3}-0.22\big|_{\mu_G^2}-0.016\big|_{\tilde{\rho}_D^3}-0.00026\big|_{\frac{1}{m_b^3}}+0.0086\big|_{\rm{IC}}-0.0018\big|_{\frac{1}{m_b^5}}\Big)
\end{align}
where we show the effect of the different orders up to $\mathcal{O}(1/m_b^5)$. Note that we also see the partial cancellation between IC and ``genuine'' $1/m_b^5$ in the total rate.
\section{Conclusion and Outlook}
\vspace*{-4mm}
\noindent
In this work, we studied the dimension-8 contributions for inclusive semileptonic $B\to X_c l\bar{\nu}$ decays. These include the Intrinsic Charm contributions proportional to $1/(m_b^3 m_b^2)$ and ``genuine'' $1/m_b^5$ contributions. Even though  power-counting arguments suggest IC contributions are required to complete the analysis at $\mathcal{O}(1/m_b^4)$, we find that the IC and ``genuine'' $1/m_b^5$ contributions enter the $E_l$ and $q^2$ with opposite sign, resulting in an overall unexpectedly small contribution from the dimension-8 operators, suggesting the IC should only be included in a combined determination up to $\mathcal{O}(1/m_b^5)$. For the $M_X^2$ moments, the IC (and total dimension-8) contributions turn out to be tiny by themselves. In the future, these $1/m_b^5$ contributions will allow for checks of the conversion of the HQE, once checked against data, and may provide a better understanding of the theory correlations that enter $|V_{cb}^{\rm{incl}}|$ determinations due to the truncation of the $1/m_b$ expansion.\vspace{-4.5mm}
\section*{Acknowledgements}
\vspace*{-4mm}
\noindent
\footnotesize{The work of I.S.M. and T.M. was supported by the Deutsche Forschungsgemeinschaft (DFG, German Research Foundation) under grant 396021762 -- TRR 257 ``Particle Physics Phenomenology after the Higgs Discovery". K.K.V. acknowledges support from the project “Beauty decays: the quest for extreme
precision” of the Open Competition Domain Science which is financed by the Dutch Research
Council (NWO).}

\scriptsize{
\bibliographystyle{JHEP}
\bibliography{refs}}

\providecommand{\href}[2]{#2}\begingroup\raggedright\begin{thebibliography}{10}

\bibitem{Bordone:2021oof}
M.~Bordone, B.~Capdevila and P.~Gambino, \emph{{Three loop calculations and inclusive Vcb}}, \href{http://dx.doi.org/10.1016/j.physletb.2021.136679}{\emph{Phys. Lett. B} {\bf 822} (2021) 136679}, [\href{http://arxiv.org/abs/2107.00604}{{\tt 2107.00604}}].

\bibitem{Bernlochner:2022ucr}
F.~Bernlochner, M.~Fael, K.~Olschewsky, E.~Persson, R.~van Tonder, K.~K. Vos et~al., \emph{{First extraction of inclusive V$_{cb}$ from q$^{2}$ moments}}, \href{http://dx.doi.org/10.1007/JHEP10(2022)068}{\emph{JHEP} {\bf 10} (2022) 068}, [\href{http://arxiv.org/abs/2205.10274}{{\tt 2205.10274}}].

\bibitem{Finauri:2023kte}
G.~Finauri and P.~Gambino, \emph{{The q$^{2}$ moments in inclusive semileptonic B decays}}, \href{http://dx.doi.org/10.1007/JHEP02(2024)206}{\emph{JHEP} {\bf 02} (2024) 206}, [\href{http://arxiv.org/abs/2310.20324}{{\tt 2310.20324}}].

\bibitem{Gambino:2020jvv}
P.~Gambino et~al., \emph{{Challenges in semileptonic $B$ decays}}, \href{http://dx.doi.org/10.1140/epjc/s10052-020-08490-x}{\emph{Eur. Phys. J. C} {\bf 80} (2020) 966}, [\href{http://arxiv.org/abs/2006.07287}{{\tt 2006.07287}}].

\bibitem{Mannel:2010wj}
T.~Mannel, S.~Turczyk and N.~Uraltsev, \emph{{Higher Order Power Corrections in Inclusive B Decays}}, \href{http://dx.doi.org/10.1007/JHEP11(2010)109}{\emph{JHEP} {\bf 11} (2010) 109}, [\href{http://arxiv.org/abs/1009.4622}{{\tt 1009.4622}}].

\bibitem{Belle:2021idw}
{\scshape Belle} collaboration, R.~van Tonder et~al., \emph{{Measurements of $q^2$ Moments of Inclusive $B \rightarrow X_c \ell^+ \nu_{\ell}$ Decays with Hadronic Tagging}}, \href{http://dx.doi.org/10.1103/PhysRevD.104.112011}{\emph{Phys. Rev. D} {\bf 104} (2021) 112011}, [\href{http://arxiv.org/abs/2109.01685}{{\tt 2109.01685}}].

\bibitem{Belle-II:2022evt}
{\scshape Belle-II} collaboration, F.~Abudin\'en et~al., \emph{{Measurement of lepton mass squared moments in B\textrightarrow{}Xc\ensuremath{\ell}\ensuremath{\nu}\textasciimacron{}\ensuremath{\ell} decays with the Belle II experiment}}, \href{http://dx.doi.org/10.1103/PhysRevD.107.072002}{\emph{Phys. Rev. D} {\bf 107} (2023) 072002}, [\href{http://arxiv.org/abs/2205.06372}{{\tt 2205.06372}}].

\bibitem{Bigi:2009ym}
I.~Bigi, T.~Mannel, S.~Turczyk and N.~Uraltsev, \emph{{The Two Roads to 'Intrinsic Charm' in B Decays}}, \href{http://dx.doi.org/10.1007/JHEP04(2010)073}{\emph{JHEP} {\bf 04} (2010) 073}, [\href{http://arxiv.org/abs/0911.3322}{{\tt 0911.3322}}].

\bibitem{Breidenbach:2008ua}
C.~Breidenbach, T.~Feldmann, T.~Mannel and S.~Turczyk, \emph{{On the Role of 'Intrinsic Charm' in Semi-Leptonic B-Meson Decays}}, \href{http://dx.doi.org/10.1103/PhysRevD.78.014022}{\emph{Phys. Rev. D} {\bf 78} (2008) 014022}, [\href{http://arxiv.org/abs/0805.0971}{{\tt 0805.0971}}].

\bibitem{Fael:2019umf}
M.~Fael, T.~Mannel and K.~K. Vos, \emph{{The Heavy Quark Expansion for Inclusive Semileptonic Charm Decays Revisited}}, \href{http://dx.doi.org/10.1007/JHEP12(2019)067}{\emph{JHEP} {\bf 12} (2019) 067}, [\href{http://arxiv.org/abs/1910.05234}{{\tt 1910.05234}}].

\bibitem{Mannel:2023yqf}
T.~Mannel, I.~S. Milutin and K.~K. Vos, \emph{{Inclusive semileptonic $ b\to c\ell \overline{\nu} $ decays to order $ 1/{m}_b^5 $}}, \href{http://dx.doi.org/10.1007/JHEP02(2024)226}{\emph{JHEP} {\bf 02} (2024) 226}, [\href{http://arxiv.org/abs/2311.12002}{{\tt 2311.12002}}].

\bibitem{Kobach:2017xkw}
A.~Kobach and S.~Pal, \emph{{Hilbert Series and Operator Basis for NRQED and NRQCD/HQET}}, \href{http://dx.doi.org/10.1016/j.physletb.2017.06.026}{\emph{Phys. Lett. B} {\bf 772} (2017) 225--231}, [\href{http://arxiv.org/abs/1704.00008}{{\tt 1704.00008}}].

\bibitem{Mannel:2018mqv}
T.~Mannel and K.~K. Vos, \emph{{Reparametrization Invariance and Partial Re-Summations of the Heavy Quark Expansion}}, \href{http://dx.doi.org/10.1007/JHEP06(2018)115}{\emph{JHEP} {\bf 06} (2018) 115}, [\href{http://arxiv.org/abs/1802.09409}{{\tt 1802.09409}}].

\bibitem{Luke:1992cs}
M.~E. Luke and A.~V. Manohar, \emph{{Reparametrization invariance constraints on heavy particle effective field theories}}, \href{http://dx.doi.org/10.1016/0370-2693(92)91786-9}{\emph{Phys. Lett. B} {\bf 286} (1992) 348--354}, [\href{http://arxiv.org/abs/hep-ph/9205228}{{\tt hep-ph/9205228}}].

\bibitem{Manohar:2010sf}
A.~V. Manohar, \emph{{Reparametrization Invariance Constraints on Inclusive Decay Spectra and Masses}}, \href{http://dx.doi.org/10.1103/PhysRevD.82.014009}{\emph{Phys. Rev. D} {\bf 82} (2010) 014009}, [\href{http://arxiv.org/abs/1005.1952}{{\tt 1005.1952}}].

\bibitem{Dugan:1991ak}
M.~J. Dugan, M.~Golden and B.~Grinstein, \emph{{On the Hilbert space of the heavy quark effective theory}}, \href{http://dx.doi.org/10.1016/0370-2693(92)90493-N}{\emph{Phys. Lett. B} {\bf 282} (1992) 142--148}.

\bibitem{Chen:1993np}
Y.-Q. Chen, \emph{{On the reparametrization invariance in heavy quark effective theory}}, \href{http://dx.doi.org/10.1016/0370-2693(93)91018-I}{\emph{Phys. Lett. B} {\bf 317} (1993) 421--427}.

\bibitem{Alberti:2014yda}
A.~Alberti, P.~Gambino, K.~J. Healey and S.~Nandi, \emph{{Precision Determination of the Cabibbo-Kobayashi-Maskawa Element $V_{cb}$}}, \href{http://dx.doi.org/10.1103/PhysRevLett.114.061802}{\emph{Phys. Rev. Lett.} {\bf 114} (2015) 061802}, [\href{http://arxiv.org/abs/1411.6560}{{\tt 1411.6560}}].

\bibitem{Gambino:2013rza}
P.~Gambino and C.~Schwanda, \emph{{Inclusive semileptonic fits, heavy quark masses, and $V_{cb}$}}, \href{http://dx.doi.org/10.1103/PhysRevD.89.014022}{\emph{Phys. Rev. D} {\bf 89} (2014) 014022}, [\href{http://arxiv.org/abs/1307.4551}{{\tt 1307.4551}}].

\bibitem{Fael:2024fkt}
M.~Fael, I.~S. Milutin and K.~K. Vos, \emph{{Kolya: an open-source package for inclusive semileptonic $B$ decays}},  \href{http://arxiv.org/abs/2409.15007}{{\tt 2409.15007}}.

\bibitem{Fael:2024rys}
M.~Fael, M.~Prim and K.~K. Vos, \emph{{Inclusive $B\rightarrow X_c \ell {\bar{\nu }}_\ell$ and $B\rightarrow X_u \ell {\bar{\nu }}_\ell$ decays: current status and future prospects}}, \href{http://dx.doi.org/10.1140/epjs/s11734-024-01090-w}{\emph{Eur. Phys. J. ST} {\bf 233} (2024) 325--346}.

\bibitem{Heinonen:2014dxa}
J.~Heinonen and T.~Mannel, \emph{{Improved Estimates for the Parameters of the Heavy Quark Expansion}}, \href{http://dx.doi.org/10.1016/j.nuclphysb.2014.09.017}{\emph{Nucl. Phys. B} {\bf 889} (2014) 46--63}, [\href{http://arxiv.org/abs/1407.4384}{{\tt 1407.4384}}].

\end{thebibliography}\endgroup

\end{document}